# Hybrid Bound States in Continuum for Enhanced Sensing and Light Manipulation

*Maik Meudt [†], Chakan Bogiadzi[†], Kevin Wrobel[†] and Patrick Görrn*[†]*

M. Meudt, C. Bogiadzi, K. Wrobel and Prof. P. Görrn

[†]University of Wuppertal, School of Electrical, Information and Media Engineering,

Chair of Large Area Optoelectronics, Rainer-Gruenter-Str. 21, 42119 Wuppertal, Germany

E-mail: goerrn@uni-wuppertal.de

# Introductory paragraph:

Light can be influenced by permittivity changes in optical resonators, enabling optical sensors, modulators and optical switches. It is straightforward that a high relative change of intensity per change of permittivity, labelled as figure of merit FOM*, is sought. This FOM* is proportional to the product of quality factor $Q$ and sensitivity $S$ of the resonator. In known resonators, an increase of $Q$ is always accompanied by a decrease of $S$ leaving FOM* constant. Hybridization of resonators has always been reported to lead to an averaging of their performance, only.
Here, we theoretically show that light diffracted by bound states in continuum (BICs) breaks that rule. Its FOM* is strongly increased by hybridization, thus outperforming both purely dielectric or plasmonic BICs. We suggest a symmetric waveguide geometry for realising topologically protected hybrid BICs, develop a polymer based fabrication technology and show first experimental evidence of hybrid BICs.

# Introduction

Optical resonators have been established as the most advanced platform for optical sensors and modulators, and they come in a broad variety of different resonator types such as localised and delocalised plasmons, photonic crystal cavities, guided mode resonances, ring resonators, and dielectric whispering gallery mode resonators[1–11].

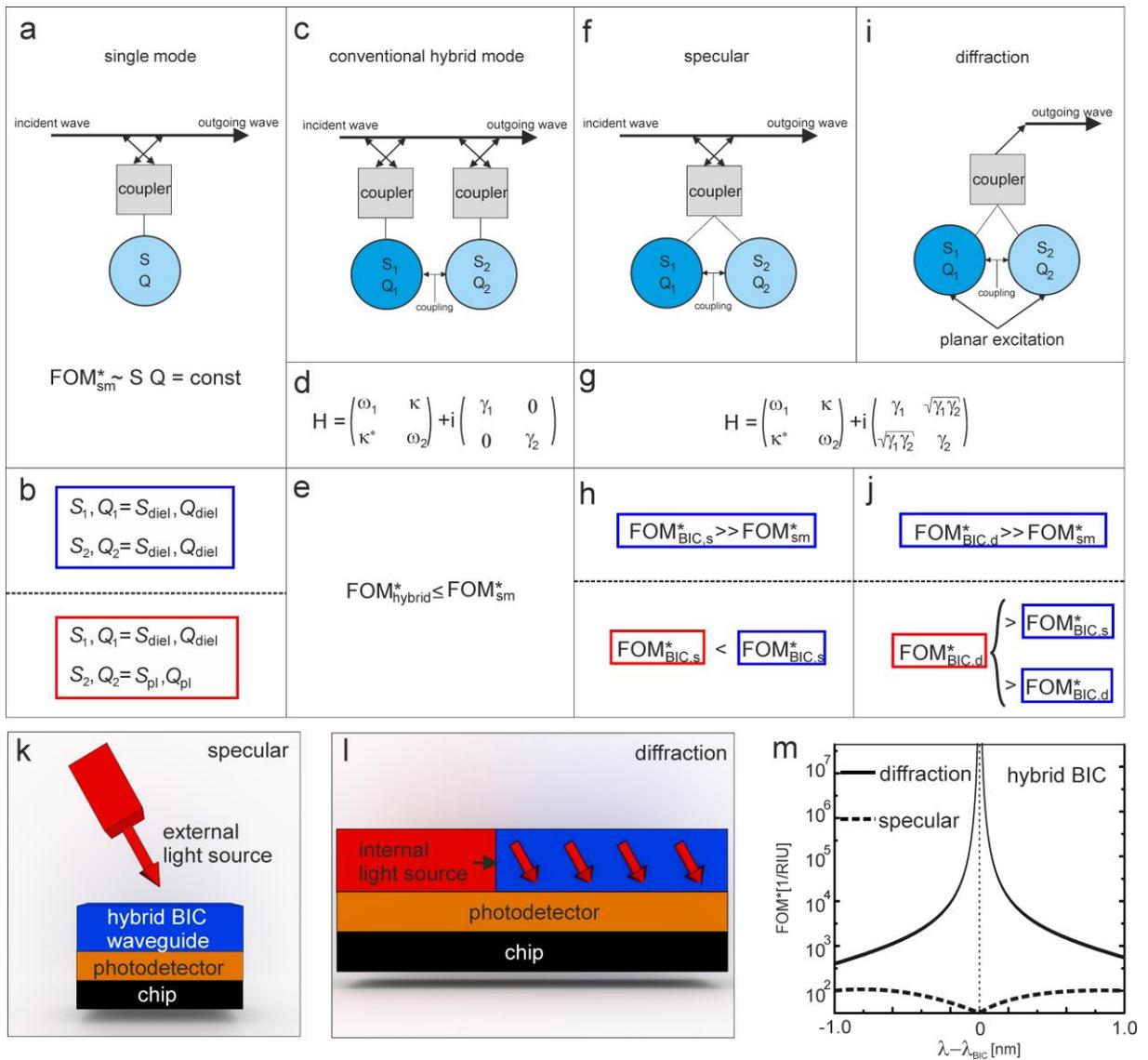

Figure 1. Comparison of conventional resonators and bound states in continuum. a) Single mode resonators exhibit a trade-off between their $Q$ factor and sensitivity $S$ and thus show a limited relative intensity change per refractive index change, labelled as $\text{FOM}^*_{sm}$. b) Combinations of resonators considered for coupled systems. Blue frame: Two lossless dielectric resonators with low $S$ and high $Q$. Red frame: One lossless dielectric resonator with low $S$ and high $Q$, and one plasmonic resonator with high $S$ and low $Q$. c) Hybrid resonator

consisting of two coupled modes and its corresponding coupled mode Hamiltonian (d)). e) The resulting hybrid modes cannot enhance the performance beyond the performance of the individual uncoupled modes. f) Coupled resonators supporting bound states in continuum. The corresponding Hamiltonian (g)) involves a complex valued coupling constant which leads to a diverging radiative $Q$ factor at a distinct frequency. h) The corresponding performance surpasses the performance of conventional resonators. However, losses lead to a strong reduction of the FOM* and hinder the utilization of plasmonics (bottom frame). i) Diffraction mode of a BIC. The BIC diffracts its power from an excited state into free space radiation. j) The corresponding FOM* benefits from the large sensitivity of surface plasmons although loss is present. k,l) Possible realisation of hybrid BICs emphasising its potential for on-chip integration. m) Corresponding FOM*. The diffraction mode hybrid BIC shows a divergent FOM* while the specular mode shows a strongly reduced FOM*.

Their performance is commonly evaluated by a figure of merit (FOM*) which is given by the relative change of a measured intensity or power signal

$$FOM^* = \frac{1}{P}\frac{\partial P}{\partial n} \propto QS.$$

For achieving high performance, high quality factors $Q = \lambda/\Delta\lambda$ and a large shift of the central resonance wavelength with respect to permittivity changes, labelled as sensitivity $S = \partial\lambda/\partial n$, are desired at the same time.

However, dielectric systems tend to have high values of $Q$ and low values of $S$ while plasmonic systems behave vice versa, limiting the achievable FOM*. It has often been considered to circumvent this limit by hybridising dielectric and plasmonic resonators[9,12–22]. For the following explanations, we symbolize a system of two dielectric resonators by a blue frame and a hybrid system of one dielectric resonator and one plasmonic resonator by a red frame as shown in Fig. 1b. An exemplary hybrid system is visualised in Fig. 1c. In the simplest case of two coupled resonators, this coupling can be expressed by a coupled mode Hamiltonian with the coupling constant $\kappa$ as shown in Fig. 1d. In fact, the resulting hybrid resonators exhibit mediocre values of $Q$ and $S$ (Fig. 1e). The same behaviour is present for the various theoretical and experimental examples cited above. Although hybrid resonators still resemble a large ongoing field of research due to their benefits of cost efficiency, none of these hybrid resonators have shown superior performance in comparison to the individual uncoupled resonators yet.

Remarkably, dielectric bound states in continuum (BICs) exhibit a higher FOM* than conventional optical resonators of the same $QS$[23,24]. This is a consequence of the destructive interference of at least two waves with a common output channel, as it is schematically shown in Fig. 1f and mathematically described in Fig. 1g[23]. The radiative component of $Q$ diverges at an exceptional frequency, while $S$ remains unaffected, leading to a strongly increased FOM* (Fig. 1h). Because of that, BICs provide high performance which has been utilised for lasers[23,25,26], sensors[27] and filters[28]. However, the FOM* of BICs remains proportional to $QS$. As the BIC combines two modes one can ask how the $QS$ of the BIC is related to the $QS$ of each mode. Historically BICs are characterized spectroscopically, thus measuring the transmission and reflection. The $QS$ of such specular BICs is below the average of the two involved modes. That is due to the fact that the sensitivity is the arithmetic mean of both modes' sensitivity, while the quality factor is the harmonic mean of $Q_1$ and $Q_2$. The hybridization results thus in a quality factor that is lower than the smaller involved quality factor. Hence, one could conclude that – like in other hybrids – BICs cannot benefit from the combination of largely different modes (Fig. 1h).

In fact, modern photonic devices have been evolving towards on-chip integration with internal light sources, photodetectors and modulators as they allow substantially higher integration density and utilise the outcoupled diffracted power from propagating modes as schematically visualised in Fig. 1i[18,29–32]. We analysed the Hamiltonian and the corresponding FOM* in Fig. 1e in detail with respect to this outcoupled diffracted power, for which the relative radiative quality factor is crucial (see supporting information). Different from all other known systems this FOM* increases when modes of different properties interfere. While the sensitivity is again an average of both sensitivities, the relative radiative quality factor is close to the one for the lossless case.

For instance, in diffraction mode the FOM* of a hybrid BIC combining a dielectric mode (high $Q_D$, low $S_D$) and a plasmon mode (low $Q_p$, high $S_p$) possesses a relative radiative quality factor close to a purely dielectric BIC. That means the ohmic losses of the plasmon do not lower the relative radiative quality factor. On the other hand, the sensitivity is largely enhanced through the contribution of the plasmon. (Fig. 1j). This advantage becomes clear for the comparison of a specular (Fig. 1k) and diffraction (Fig. 1l) setup utilising hybrid BICs. The FOM* of these two systems is shown in Fig. 1m. While the specular system (dashed line) exhibits decreasing values towards the wavelength of the hybrid BIC, the values for the diffraction system (black line) diverge to infinity.

In summary, the theoretical analysis predicts that hybrid BICs integrated in the diffraction mode will enable the highest possible FOM*. However, so far an experimental realisation of hybrid BICs has not been reported. Most recently, Azzam et. al have theoretically predicted bound states in continuum in hybrid waveguides[33]. Surprisingly, no further exploration of their performance has been considered, yet.

Here we design a hybrid waveguide of certain symmetry for which topologically protected hybrid BICs exist, and we theoretically show that its FOM* surpasses the performance of purely dielectric systems as a result of the enhanced sensitivity by surface plasmons. We further provide a strategy on how to fabricate the otherwise technically challenging fully symmetric hybrid waveguides by cost-efficient methods and show experimental evidence for the existence of the proposed hybrid BICs.

## Results & Discussion

The proposed hybrid waveguide consists of a layer of OrmoCore with a thin silver film in its exact centre. This design is visualised in Fig. 2a. Its symmetry ensures the topological protection of the hybrid BICs as it is shown later on. At the energy range of interest, the hybrid waveguide supports two propagating photonic modes and two plasmonic modes respectively as it is indicated by the dispersion relation in Fig. 2b. Because the hybrid waveguide is mirror symmetric, the magnetic field distributions shown in Fig. 2c are either symmetric or antisymmetric with respect to the z=0 plane. The symmetric and antisymmetric photonic modes with respective $Q$-factors of 5560 and 2120 and rather low values of $S$ around 40 and 80 nm/RIU are labelled as $PH_s$ and $PH_a$, while the plasmonic modes possess low $Q$-factors of 535 and 160 combined with large sensitivies around 500 and 700 nm/RIU and are termed symmetric surface plasmon polariton (SPP$_s$) and antisymmetric surface plasmon polariton (SPP$_a$)[34].

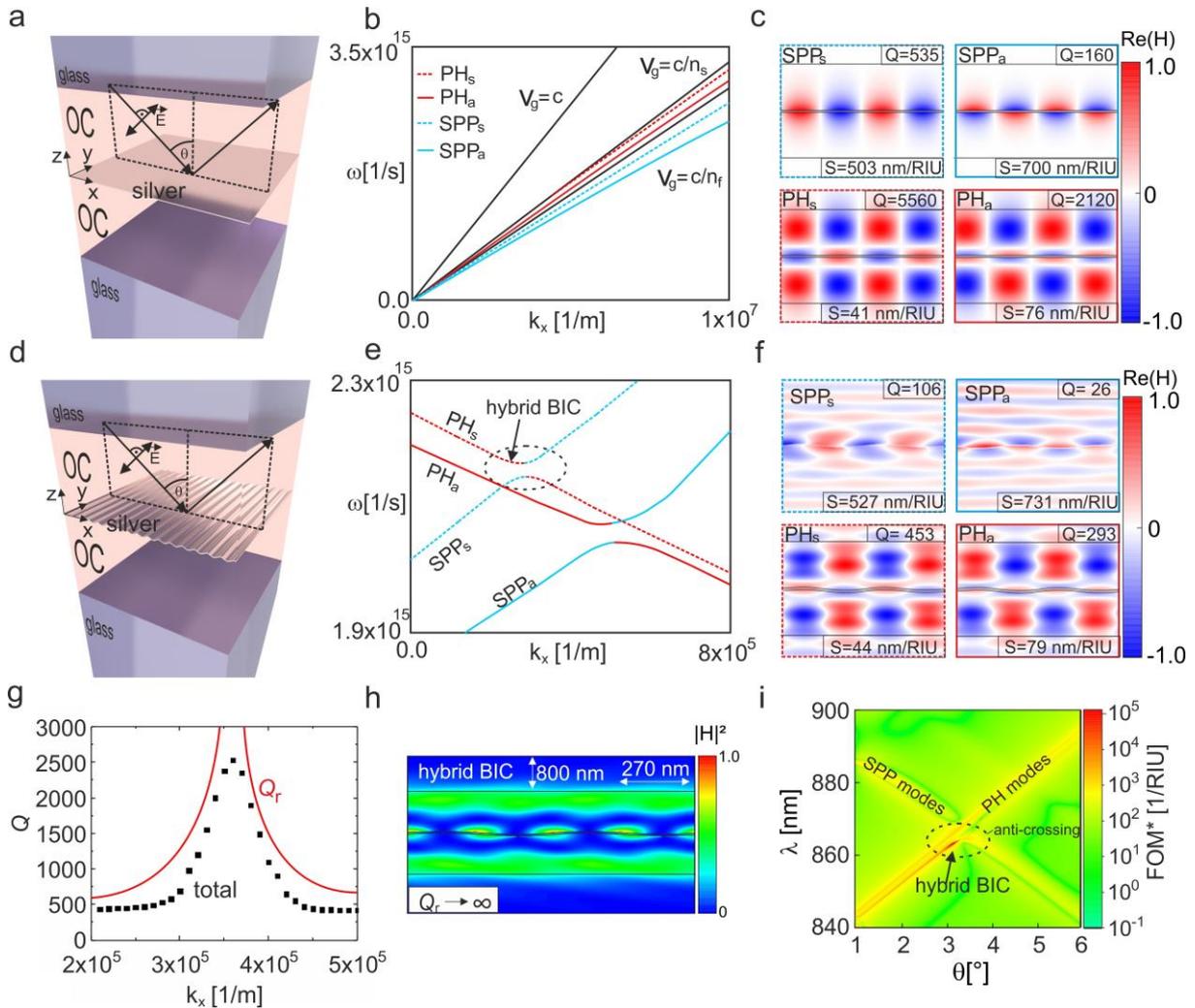

Figure 2. a) Hybrid waveguide consisting of a flat silver film embedded in an OrmoCore (OC) waveguide with glass cladding. b) Dispersion relation of the hybrid waveguide in a). c) Cross-sectional plots of Re (H) of the symmetric plasmonic mode (SPP$_s$), the antisymmetric plasmonic mode (SPP$_a$), the symmetric photonic mode (PH$_s$) and the antisymmetric photonic mode (PH$_a$) with corresponding $Q$-factors and sensitivities. d) Modified version of a) with a sinusoidally modulated silver film. e) Dispersion relation of the hybrid waveguide in d) with the anti-crossing of modes. f) Cross sectional field plots of Re (H) of the SPP$_s$, SPP$_a$, PH$_s$ and PH$_a$ mode, $Q$-factors and sensitivities. g) Plot of the total quality factor (black dots) and the radiative quality factor $Q_r$ of the upper PH$_s$-SPP$_s$ dispersion line as function of k$_x$ (see ellipsoid in e). h) Magnetic field intensity |H|² of the hybrid BIC. i) Simulated FOM* of the hybrid waveguide. Towards the hybrid BIC, the FOM* strongly increases by more than two orders of magnitude under realistic assumptions of deviations from an ideal geometry (see methods).

In order to achieve hybrid bound states in continuum, the SPP modes and PH modes have to be coupled with a complex valued coupling constant. This is accomplished by a sinusoidal modulation of the silver film as indicated by Fig. 2d, which leads to the formation of an optical band structure[35]. Additionally, such a modulation provides time reversal symmetry as well as mirror symmetry under glide operations, which is described by the expression

$$\varepsilon(x + \delta, y, -z) = \varepsilon^*(x, y, z).$$

Together with the symmetric dielectric environment, this ensures topologically protected BICs because of equal Fourier amplitudes in both the z and –z direction[24,36]. Consequently, changes of the permittivity or geometry lead to a shift of the BIC position in the energy momentum space only, without affecting its stability.

The coupling between the modes manifests as an anti-crossing of their dispersion curves in the resulting band structure of the hybrid waveguide (Fig. 2e). Spectrally far away from the position of anti-crossing, the diffraction into free space radiation becomes visible as strong periodic distortions of the cross-sectional field amplitudes, as shown in Fig. 2f. These distortions can be understood as sources for the formation of plane waves emitting power into the radiation continuum. They build a common output channel for both the SPP modes and PH modes. For the majority of the calculated frequencies and momenta, the emission of power into this output channel causes substantially lower $Q$-factors than for the undisturbed modes of the planar geometry in Fig. 1c. However, towards a distinct frequency and momentum near the position of anti-crossing, the $Q$-factor along the $PH_s$-dispersion curve visualised in Fig. 2g exhibits a strongly pronounced maximum of a value of 2520. We observe that this maximum occurs as the result of the divergence of the radiative component of the $Q$-factor, labelled as $Q_r$. In accordance with this, the corresponding cross-sectional field amplitude at this distinct frequency in Fig. 2h shows an undisturbed evanescent field decay in the waveguide claddings, independently proving that losses due to radiation into free space modes are completely suppressed by destructive interference between the $SPP_s$ mode and $PH_s$ mode. This confirms that a hybrid BIC is indeed formed by the coupling of a photonic high $Q$ mode and a plasmonic high $S$ mode in the proposed hybrid waveguide.

In order to emulate the conditions of an on-chip system with an internal light source, we assume that a propagating mode on the dispersion line of the PH$_s$ mode near the distinct frequency of the hybrid BIC is excited with an initial power of $P_{in}$ as emphasised in Fig. 1l. We analyse its outcoupled diffracted power. As shown in the supporting information, the FOM* with respect to that outcoupled diffracted power can be approximated by the relative change of the radiative extinction rate

$$FOM^* = \frac{1}{\gamma_r}\frac{\partial \gamma_r}{\partial n}$$

and is visualised in Fig. 2h, whereby the radiative extinction rate follows the proportionality $\gamma_r \propto 1/Q_r$. Please note, that for a better comparison with the simulation results and experimental results later on, the FOM* is plotted against the light source wavelength $\lambda$ and outcoupling angle $\theta$ instead of the frequency and momentum. First of all, the FOM* is close to zero for all wavelengths and angles which do not fall together with the dispersion curves of the modes. This is not surprising as no guided mode solutions exist there. Along the dispersion curves of the modes, the FOM* possesses somewhat higher values of about 200/RIU for the SPP modes and around 1000/RIU for the PH modes for the majority of the displayed wavelengths and angles. Most strikingly, towards the wavelength and outcoupling angle of the hybrid BIC at 860 nm and 3°, we observe that the FOM* strongly increases to a value of more than $10^5$/RIU and even surpasses the values which would be expected for purely dielectric BICs by almost one order of magnitude[37]. This is because the relative radiative Q-factor does not decrease in comparison to the purely dielectric case despite the presence of losses (see supporting information). At the same time, the presence of surface plasmons increases the average sensitivity $\bar{S}$, leading to an approximate relation

$$FOM^*_{\text{hybrid}} \approx FOM^*_{\text{diel}} \cdot \frac{\bar{S}}{S_{\text{diel}}} \approx 8 \cdot FOM^*_{\text{diel}}$$

with $S_{\text{diel}}$ as the sensitivity of the PH$_s$ mode.

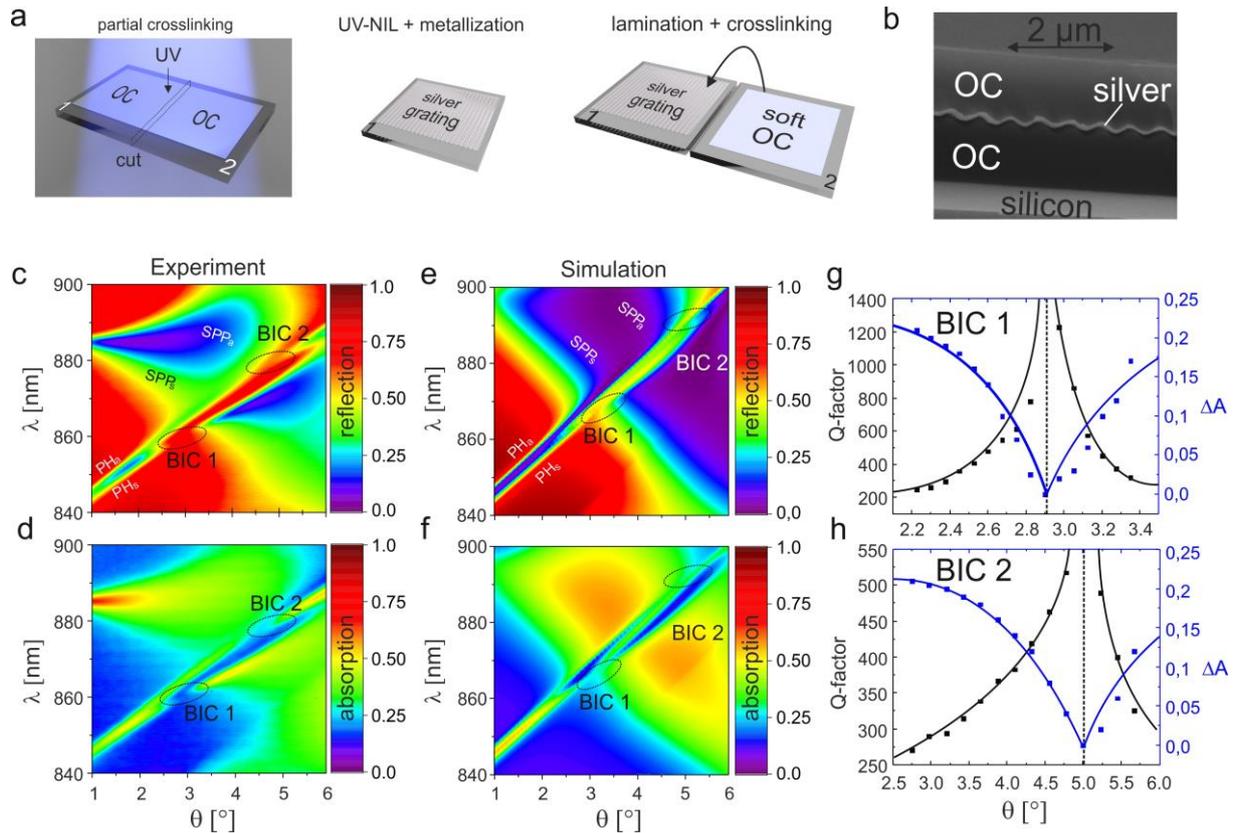

Figure 3. a) Fabrication of symmetric hybrid waveguides. b) Cross-sectional SEM micrograph of the hybrid waveguide. c,d) Measured reflection and absorption spectra of a hybrid waveguide laminated between two glass slides. Two BICs are observed at 2.9° and 860 nm as well as 5° and 880 nm. e,f) Simulated reflection and absorption spectra of an ideal hybrid waveguide. g,h) Quality factors and amplitudes extracted from the fits of a Lorentzian model to the measured absorption spectra.

This enhanced FOM* of hybrid BICs appears promising for future high performance applications. However, the realisation of such symmetric hybrid waveguides is technologically challenging. It requires the fabrication of metallic volume structures perfectly centred in a photonic waveguide. We identified polymer technology and mechanical structuring methods to be perfectly suited for this purpose. In addition, polymer technology comes along with several advantages, which are emphasised with the help of Fig. 3a. OrmoCore is crosslinkable by UV irradiation. Its Young's Modulus can be controlled by the applied UV dose. We use this property to fabricate a hardened waveguide with a nanoimprinted and subsequently metallised volume grating as well as a waveguide with a mechanically soft and equally thick OrmoCore layer. When both waveguides are mechanically pressed together, the soft OrmoCore layer adapts to the curvature of the volume grating and then hardens under a final UV irradiation step, resulting in a symmetric hybrid waveguide. All these steps are upscalable, provide perfect centering of the silver film, and are even suited for achieving free-standing waveguides. Further details on the fabrication procedure are given in the methods section.

Following this procedure, we successfully fabricated a symmetric hybrid waveguide. A scanning electron micrograph (SEM) of the cross section shown in Fig. 3b confirms the symmetry of the waveguide as well as the sinusoidal shape of the silver film.

For its optical characterisation we investigated the hybrid waveguide by angle resolved reflection and transmission spectroscopy. The resulting optical transmission and reflection spectra of the hybrid waveguide (Fig. 3c and d) show spectrally distinct Fano shaped resonances, which vary with differing incident angles. These resonances can be identified as the presumed PH modes and SPP modes and are labelled in the corresponding graphs. A comparison with the spectra of a simulated hybrid waveguide with the same geometrical parameters displayed in Fig. 3e and f shows overall qualitative agreement, yet differs quantitatively with respect to the plasmon modes. This is most likely due to slight variations from the idealised sinusoidal shape assumed in the simulation. Remarkably, the absorption peaks of the $PH_s$ and $PH_a$ dispersion curves disappear near the predicted wavelengths and angles. The analysis of the resonances around these exceptional points, shown in Fig. 3g and h, reveals a strong increase of the Q-factor as well as a decrease of the absorption amplitudes to zero, which matches the theoretical predictions in the direct vicinity of a hybrid BIC. Together, these observations confirm the existence of hybrid BICs.

The fact that these states could be experimentally observed, despite slight deviations of the plasmon resonances from the ideal simulated case, is further interpreted as a result of their topological protection. Most importantly, the excellent agreement of the Q-factor and absorption amplitude between the experiment and simulation lead to the conclusion that the observed hybrid BICs behave as predicted in Fig. 2.

From a broader perspective, the ability to produce hybrid waveguides of high symmetry opens up new opportunities for BICs in general and enables the production of transfer printable free standing

BIC waveguides, their combination with silicon substrates, 3D stacking for the realisation of both complex and cost efficient high FOM* sensors and modulators. By using plasmonic materials with low plasmon frequencies, it is straightforward to adapt hybrid BICs to the telecommunication window[38]. We further anticipate hybrid BICs for logical components with low energy consumption and ultra-sensitive photonic devices with enhanced light-matter interaction utilising active, non-linear materials.

In summary, we have theoretically investigated hybrid bound states in continuum in symmetric hybrid waveguides and experimentally demonstrated their existence. Arising from the coupling between long range surface plasmon polaritons and photonic waveguide modes, these states combine the diverging radiative quality factor and an increased sensitivity due to strong plasmonic field confinement. Hybrid BICs in diffraction mode enable substantially better performance than dielectric BICs despite the introduction of a lossy material and, in contrast to conventional hybrid resonators, show a new direction towards the long-sought idea of enhancing photonics with hybrid systems[21,39]. As the total quality factor in direction of propagation does not diverge, the hybrid BIC can even be observed for systems of limited size, thus suggest higher integration densities than for existing BIC systems, and show higher values of the FOM* than previously reported.

## Methods

### Simulation

All simulations shown in this letter have been calculated by standard rigorous coupled wave analysis (RCWA).

The optical constants used for glass and OrmoCore were calculated according to the Cauchy-model

$$n(\lambda) = A + \frac{B}{\lambda^2} + \frac{C}{\lambda^4}$$

with $A_{glass} = 1.4613, B_{glass} = 0.00299$ µm² and $A_{OC} = 1.533, B_{OC} = 0.00617$ µm². The optical constants of silver were taken from Palik et. al.

The plane of incidence was defined in the x-z-plane with p-polarised light, whereas the geometrical parameters in the simulation model were chosen to $t_{core} = 2130$ nm, $t_{grating} = 100$ nm, $t_{silver} = 60$ nm and $\Lambda = 555$ nm (see Fig S5a). For the calculation of the FOM* in Fig 2i, deviations from an ideal geometry (for example by fabrication imperfections) were assumed to cause a parasitic radiative scattering rate of $\gamma_{sca}/\gamma_r$=0.01.

### Sample Fabrication

The approach to achieving free-standing hybrid waveguides is shown in Fig. S5b. First, a glass substrate was coated by a functional layer of water-soluble polyvinyl alcohol (PVA) and a homogeneous layer of OrmoCore. The OrmoCore film was partially crosslinked at an exposure dose around 10% of the full crosslinking dose. Then, the whole stack was separated into two pieces. Using a commercially available sinusoidal stamp, the first piece was structured by UV-nanoimprint lithography, and subsequently metallised by physical vapour deposition. The OrmoCore film on the second piece was delaminated by a lift-off in water and transferred sunny-side down onto the first piece with the help of an anti-sticking layer (ASL) coated silicon wafer[40]. Finally, the stack was fully cross-linked by UV exposure and removed from the substrate by a lift-off in water (device type A) or supported by a planar substrate and superstrate such as two glass slides (device type B).

Angular resolved reflection and transmission measurements

A sketch of the experimental setup for the optical characterisation of the hybrid waveguide is shown in Fig. S5c. A collimated helium-deuterium white light source was used as illumination signal. To obtain angle resolved measurements, both the sample and a grating spectrometer with a resolution of $\Delta\lambda = 0.01$ nm were placed in a movable position on a 2-axis rotational stage independently from each other. All spectra were measured with an angular resolution of $\Delta\theta = 0.075°$.

Q-factor extraction

The Q-factors from Fig. 3e and f were extracted by fitting a Lorentz model to the absorption spectra, which is described by

$$A(\omega) = A_0 + \Delta A \frac{\gamma^2}{(\omega^2 - \omega_0^2)^2 + \gamma^2},$$

where $A_0(\omega)$ is a slowly varying continuous background and $\Delta A$ is the absorption amplitude. $\gamma$ defines the FWHM of the resonance and is linked to the Q-factor by Q=$\omega_0/\gamma$.

**References**

1.  Petryayeva, E. & Krull, U. J. Localized surface plasmon resonance: Nanostructures, bioassays and biosensing—A review. *Anal. Chim. Acta* **706,** 8–24 (2011).

2.  Willets, K.A., Van Duyne, R.P. Localized surface plasmon resonance spectroscopy and sensing. *Annu. Rev. Phys. Chem.* **58,** 267 (2007).

3.  Mayer, K. M. & Hafner, J. H. Localized Surface Plasmon Resonance Sensors. *Chem. Rev.* **111,** 3828–3857 (2011).

4.  Homola, J., Yee, S. S. & Gauglitz, G. Surface plasmon resonance sensors: review. *Sensors Actuators B Chem.* **54,** 3–15 (1999).

5.  Chow, E., Grot, A., Mirkarimi, L. W., Sigalas, M. & Girolami, G. Ultracompact biochemical sensor built with two-dimensional photonic crystal microcavity. *Opt. Lett.* **29,** 1093–1095 (2004).

6.  Akahane, Y., Asano, T., Song, B.-S. & Noda, S. High-Q photonic nanocavity in a two-dimensional photonic crystal. *Nature* **425,** 944–947 (2003).

7.  Wang, S. S. & Magnusson, R. Theory and applications of guided-mode resonance filters. *Appl. Opt.* **32,** 2606–2613 (1993).

8.  Ksendzov, A. & Lin, Y. Integrated optics ring-resonator sensors for protein detection. *Opt. Lett.* **30,** 3344–3346 (2005).

9.  Ciminelli, C., Campanella, C. M., Dell'Olio, F., Campanella, C. E. & Armenise, M. N. Label-free optical resonant sensors for biochemical applications. *Prog. Quantum Electron.* **37,** 51–107 (2013).

10. Label-free detection with high-Q microcavities: a review of biosensing mechanisms for integrated devices. *Nanophotonics* **1,** 267 (2012).

11. Luchansky, M. S. & Bailey, R. C. High-Q Optical Sensors for Chemical and Biological

Analysis. *Anal. Chem.* **84,** 793–821 (2012).

12. Chen, X., Zhou, K., Zhang, L. & Bennion, I. Simultaneous measurement of temperature and external refractive index by use of ahybrid grating in D fiber with enhanced sensitivity by HF etching. *Appl. Opt.* **44,** 178–182 (2005).

13. Liu, Z. *et al.* Enhancing refractive index sensing capability with hybrid plasmonic–photonic absorbers. *J. Mater. Chem. C* **3,** 4222–4226 (2015).

14. Rifat, A. A. *et al.* Highly sensitive multi-core flat fiber surface plasmon resonance refractive index sensor. *Opt. Express* **24,** 2485–2495 (2016).

15. Fan, B. *et al.* Refractive index sensor based on hybrid coupler with short-range surface plasmon polariton and dielectric waveguide. *Appl. Phys. Lett.* **100,** 111108 (2012).

16. Velichko, E. A. & Nosich, A. I. Refractive-index sensitivities of hybrid surface-plasmon resonances for a core-shell circular silver nanotube sensor. *Opt. Lett.* **38,** 4978–4981 (2013).

17. Bahrami, F., Maisonneuve, M., Meunier, M., Aitchison, J. S. & Mojahedi, M. An improved refractive index sensor based on genetic optimization of plasmon waveguide resonance. *Opt. Express* **21,** 20863–20872 (2013).

18. Haffner, C. *et al.* Low-loss plasmon-assisted electro-optic modulator. *Nature* **556,** 483–486 (2018).

19. Zhou, L., Sun, X., Li, X. & Chen, J. Miniature microring resonator sensor based on a hybrid plasmonic waveguide. *Sensors (Basel).* **11,** 6856–6867 (2011).

20. Singh, S., Mishra, S. K. & Gupta, B. D. Sensitivity enhancement of a surface plasmon resonance based fibre optic refractive index sensor utilizing an additional layer of oxides. *Sensors Actuators A Phys.* **193,** 136–140 (2013).

21. Oulton, R. F., Sorger, V. J., Genov, D. A., Pile, D. F. P. & Zhang, X. A hybrid plasmonic waveguide for subwavelength confinement and long-range propagation. *Nat. Photonics* **2,** 496 (2008).

22. Zentgraf, T., Zhang, S., Oulton, R. F. & Zhang, X. Ultranarrow coupling-induced transparency bands in hybrid plasmonic systems. *Phys. Rev. B* **80,** 195415 (2009).

23. Hsu, C. W., Zhen, B., Stone, A. D., Joannopoulos, J. D. & Soljačić, M. Bound states in the continuum. *Nat. Rev. Mater.* **1,** 16048 (2016).

24. Hsu, C. W. *et al.* Observation of trapped light within the radiation continuum. *Nature* **499,** 188 (2013).

25. Rybin, M. & Kivshar, Y. Supercavity lasing. *Nature* **541,** 164 (2017).

26. Kodigala, A. *et al.* Lasing action from photonic bound states in continuum. *Nature* **541,** 196 (2017).

27. Liu, Y., Zhou, W. & Sun, Y. Optical Refractive Index Sensing Based on High-Q Bound States in the Continuum in Free-Space Coupled Photonic Crystal Slabs. *Sensors* **17,** (2017).

28. Foley, J. M., Young, S. M. & Phillips, J. D. Symmetry-protected mode coupling near normal incidence for narrow-band transmission filtering in a dielectric grating. *Phys. Rev. B* **89,** 165111 (2014).

29. Sorger, V. J., Oulton, R. F., Ma, R.-M. & Zhang, X. Toward integrated plasmonic circuits. *MRS Bull.* **37,** 728–738 (2012).

30. Fang, Y. & Sun, M. Nanoplasmonic waveguides: towards applications in integrated nanophotonic circuits. *Light Sci. Appl.* **4,** e294–e294 (2015).

31. Wang, J., Sciarrino, F., Laing, A. & Thompson, M. G. Integrated photonic quantum technologies. *Nat. Photonics* (2019). doi:10.1038/s41566-019-0532-1

32. Spencer, D. T. *et al.* An optical-frequency synthesizer using integrated photonics. *Nature* **557,** 81–85 (2018).

33. Azzam, S. I., Shalaev, V. M., Boltasseva, A. & Kildishev, A. V. Formation of Bound States in the Continuum in Hybrid Plasmonic-Photonic Systems. *Phys. Rev. Lett.* **121,** 253901 (2018).

34. Berini, P. Long-range surface plasmon polaritons. *Adv. Opt. Photonics* **1,** 484–588 (2009).

35. Joannopoulos, J. D., Villeneuve, P. R. & Fan, S. Photonic crystals. *Solid State Commun.* **102,** 165–173 (1997).

36. Bulgakov, E. N., Maksimov, D. N., Semina, P. N. & Skorobogatov, S. A. Propagating bound states in the continuum in dielectric gratings. *J. Opt. Soc. Am. B* **35,** 1218–1222 (2018).

37. Liu, Y., Wang, S., Zhao, D., Zhou, W. & Sun, Y. High quality factor photonic crystal filter at k≈0 and its application for refractive index sensing. *Opt. Express* **25,** 10536–10545 (2017).

38. Noginov, M. A. *et al.* Transparent conductive oxides: Plasmonic materials for telecom wavelengths. *Appl. Phys. Lett.* **99,** 21101 (2011).

39. Alam, M. Z., Aitchison, J. S. & Mojahedi, M. A marriage of convenience: Hybridization of surface plasmon and dielectric waveguide modes. *Laser Photon. Rev.* **8,** 394–408 (2014).

40. Thomas, M., P. De Boer, M., D. Shinn, N., J. Clews, P. & A. MICHALSKE, T. Chemical Vapor Deposition of Fluoroalkylsilane Monolayer Films for Adhesion Control in Microelectromechanical Systems. Journal of Vacuum Science & Technology B: Microelectronics and Nanometer Structure*s* **18,** (2000).

**Acknowledgements**

This project has received funding from the European Research Council (ERC) under the European Union's Horizon 2020 research and innovation programme (grant agreement No. 637367).

**Author Contributions**

The manuscript was written through contributions of all authors. All authors have given approval to the final version of the manuscript.

**Financial Competing Interests**

We hereby declare that we, the authors, have no competing interests as defined by Nature Research, or interests that might be perceived to influence the interpretation of the article.

The signed form "Disclosure of Potential Competing Interest" is attached to this letter.